\documentclass[aps,prl,twocolumn,groupedaddress,showpacs]{revtex4}
\usepackage{graphics,graphicx}
\usepackage{color}
\usepackage{amssymb,amsmath}
\usepackage{bookmath}
\begin{document}
\title{Broken translational and time-reversal symmetry in superconducting films}
\author{A. B. Vorontsov}\altaffiliation{
Present address: Dept. of Physics, Montana State University, Bozeman, MT 59717}
\affiliation{Department of Physics and Astronomy,
             Louisiana State University, Baton Rouge, Louisiana, 70803, USA}
\date{\today}

\begin{abstract}
We demonstrate that films of unconventional ($d$-wave) superconductor 
at temperatures $T\lesssim 0.43 \, T_c$ can exhibit 
unusual superconducting phases. 
The new ground states beside the broken gauge 
and the point group symmetries can spontaneously break  
(i) continuous translational symmetry and form periodic 
order parameter structures in the plane of the film, or    
(ii) time-reversal symmetry and develop supercurrent 
flowing along the film. 
These states are result of the strong 
transverse inhomogeniety present in films with 
thickness of several coherence lengths. 
We show a natural similarity between formation of these states and 
the Fulde-Ferrell-Larkin-Ovchinnikov state. 
\end{abstract}
\maketitle

\paragraph{Introduction.} 
Properties of superconducting state in confined geometry  
will have defining influence on minimization of superconducting devices. 
This is especially true of an unconventinal superfluid with 
order parameter (OP) that breaks  
more than one of the normal state symmetries\cite{vol85,sig91,vol90} 
and which can be suppressed by interfaces, 
forming new non-uniform ground states.\cite{kas00} 
Such states can exhibit new broken symmetries, 
and currently are subject of broad theoretical and experimental 
investigation.\cite{reviews} 
%
For example, they can arise in bulk superconductors 
under influence of external field or pressure. 
For example, magnetic field breaks the time-reversal symmetry by 
inducing supercurrents, or, as in case of 
Fulde-Ferrell-Larkin-Ovchinnikov (FFLO) state\cite{ful64,lar64},    
results in an oscillating OP  
that breaks continuous spatial translation symmetry.
In unconventional superconductors new 
ground states have been predicted to appear spontaneously 
even in the absence of externally applied magnetic field.
One origin of these states is the interaction of self-induced 
magnetic field, caused by distortion of the OP, 
with the OP itself.\cite{pal90,pal92,hig97,bar00a,lof00}
Other current-carrying states arise near surfaces 
due to subdominant $d+is$ pairing that breaks time-reversal symmetry.\cite{mat95,fog97a} 
Finally, new states with triplet structure may appear in quantum wires.\cite{bob04}
Ultimately, the origin of these states lies in the appearance  
of Andreev states bound to interfaces.\cite{hig97,kus99,kas00}

In this Letter we propose existence of new ground states 
with non-trivially broken symmetries that require 
neither the self-induced magnetic field 
nor complex subdominant pairing. 
These states are induced by strong distortion of the OP shape. 
We consider this problem in the context of a singlet 
$d$-wave superconductor, 
when distortion of the OP is caused 
by confinement to a film or wire. 
We assume that the thickness of the film is ajustable, 
and that by varying it we are able to drive transitions 
between different ground states. 
The simplicity of this model makes the problem more 
transparent, yet allows to draw general conclusions 
about connection of inhomogeniety and spontaneous symmetry breaking 
in unconventional pairing states, including superfluid \He.\cite{vor07a}  

\begin{figure}[b!]
\centerline{\includegraphics[width=0.99\linewidth]{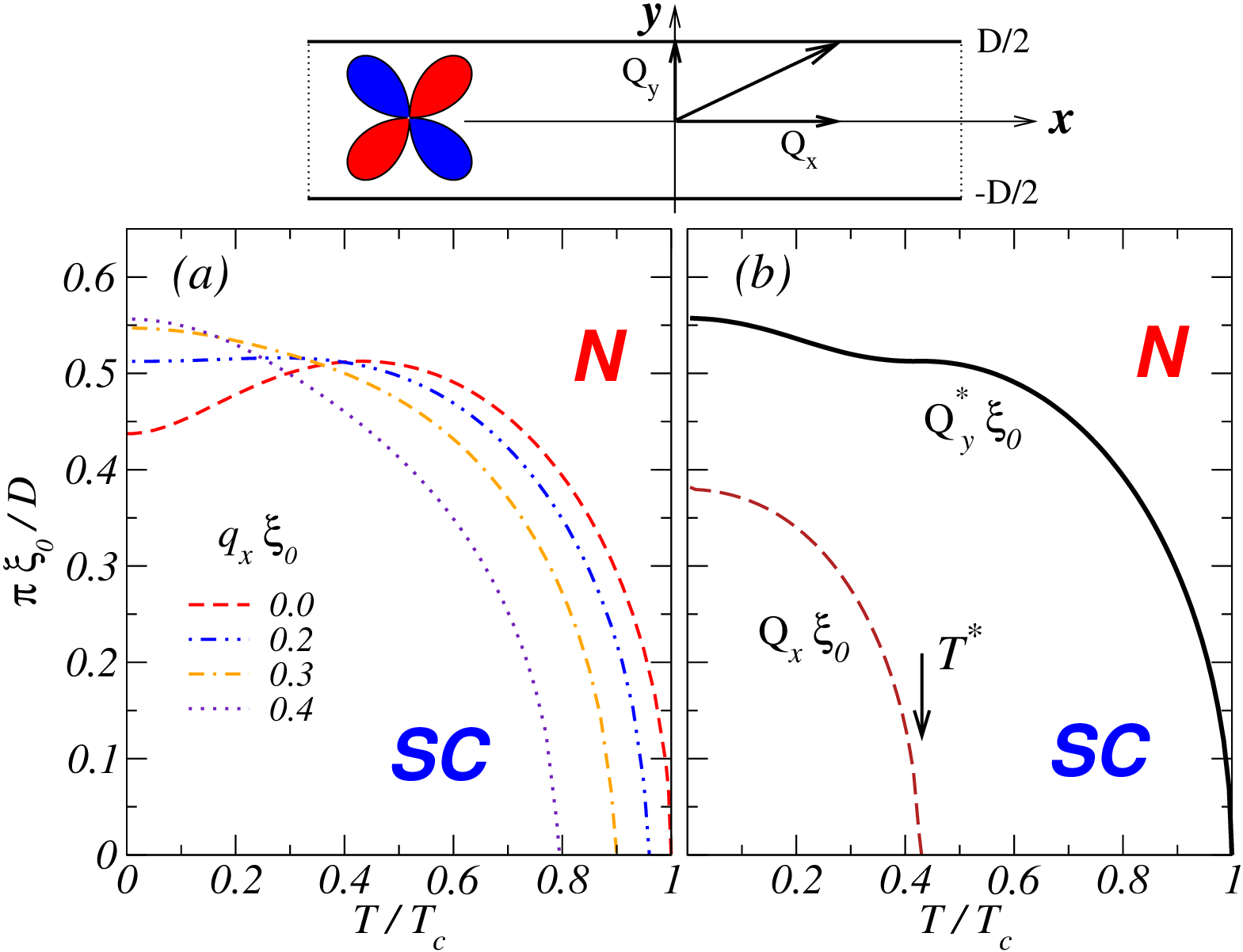}}
\caption{ (Color online) 
A $d$-wave superconductor (SC) confined to a film with pairbreaking surfaces. 
(a) 
The SC-normal(N) transition as a function of temperature 
for several $q_x$-modulated states. 
At low temperature, transition into uniform ($q_x=0$) 
SC state shows re-entrant behavior and is preempted 
by transition into a state with finite $x$-modulation; 
(b) the envelope of various $q_x$ curves gives 
the maximal confinement vector $Q^*_y(T,Q_x) = \pi/D^*$, plotted 
together with the corresponding modulation vector $Q_x(T)$. 
The coherence length $\xi_0=\hbar v_f / 2\pi T_c$. 
}
\label{fig:PDQ}
\end{figure}

\paragraph{Free energy in a film.} 
We consider a film of a $d$-wave superconductor 
with pairbreaking surfaces, such that 
the OP in the film is very non-uniform. 
This is achieved by orienting a gap node 
along the film (sketch in Fig.~\ref{fig:PDQ}). 
We assume a cylindrical Fermi surface, and OP solutions 
uniform along its axis ($z$-direction). 
Below the superconducting bulk temperature $T_c$, 
the OP can be continuously suppressed to zero 
by reducing the thickness $D$ of the film, 
eventually reaching the normal state 
below a critical value of $D^*(T)$.\cite{nag95} 
Our goal is to look for states that have more features than just the 
trivial suppression of the OP near the film's surfaces. 

We start with derivation of the Ginzburg-Landau free energy that 
describes this transition. 
For this we solve the microscopic quasiclassical equations\cite{eil68,lar68} for the 
diagonal, $g$, and off-diagonal (anomalous) components, $f$ and $f'$, 
of the quasiclassical Green's function, $\whg(\vR, \hvp; \vare_m)$. 
Complete set of equations also includes the normalization condition and symmetry relations, 
\bea
&&[\vare_m + {1\over2}\vv(\hvp) \cdot\gradR] f 
= i\,g \; \Delta(\vR, \hvp)  \,,
\label{eq:eil}
\\
&&g^2 -f f' = -\pi^2 \,, \quad
f'(\vR, \hvp; \vare_m)  = f(\vR, \hvp; -\vare_m)^* \,.
\nonumber
\eea
Here $\vv(\hvp)=v_f \hvp$ is the Fermi velocity at point $\hvp$ on the Fermi surface and 
$\vare_m= \pi T (2m+1) $ - Matsubara energy. 
We assume a separable OP, and expand the spatial part in plane waves, 
$\Delta(\vR, \hvp) =  \cY(\hvp) \sum_\vq \, \Delta_\vq e^{i\vq\vR} \,$.
The normalized basis function is $\cY(\hvp) = \sqrt{2} \sin 2\phi_{\hvp}$, 
and the plane wave amplitudes satisfy self-consistency equation, 
\be
\Delta_\vq \, \ln {T\over T_c} = T\sum_{\vare_m}  \left< \cY(\hvp) 
\left( f_\vq (\hvp; \vare_m) - {\pi \Delta_\vq \cY(\hvp) \over |\vare_m|} \right) 
\right> \,,
\label{eq:selfc}
\ee 
that follows from minimization of the free energy functional, 
$\delta \Del F / \delta \Delta^*_\vq = 0$, 
and brackets denote angle average, 
$\langle \dots \rangle = \int \done{\phi_{\hvp}} \dots $.
Solving Eqs.(\ref{eq:eil}) for $f$ up to third order in $\Delta_\vq$, 
we use self-consistency (\ref{eq:selfc}), to find the free energy,
\begin{widetext}
\begin{subequations}
\bea
&&\Del F = \sum_\vq \; I(T, \vq) \, |\Delta_\vq|^2 + 
{1\over 2} \sum_{\vq_1+\vq_2 = \vq_3+\vq_4} \, K(T, \vq_1, \vq_2, \vq_3, \vq_4) \;
\Delta_{\vq_1}^* \Delta_{\vq_2}^* \Delta_{\vq_3} \Delta_{\vq_4} \,,
\label{eq:feF}
\\
&&I(T,\vq) = \ln {T\over T_c} - 2\pi T \sum_{\vare_m>0} \Re 
\Big< \cY^2(\hvp) \left(  {1\over \vare_m + i \eta_{\vq}} - {1\over \vare_m} \right) \Big> \;,
\quad \mbox{where} \qquad \eta_\vq = \onehalf \vv \cdot \vq \,,
\label{eq:feJ}
\\
&&K(T, \vq_1, \vq_2, \vq_3, \vq_4) = 2\pi T \sum_{\vare_m>0} {1\over2} \Re 
\Big< 
\cY^4(\hvp) \frac{\vare_m+i(\eta_{\vq_1}+\eta_{\vq_2}+\eta_{\vq_3}+\eta_{\vq_4})/4}
{(\vare_m+i\eta_{\vq_1}) (\vare_m+i\eta_{\vq_2}) (\vare_m+i\eta_{\vq_3}) (\vare_m+i\eta_{\vq_4})}
\Big> \,.
\label{eq:feK}
\eea
\label{eq:fe}
\end{subequations}
\end{widetext}

This functional describes second-order transition from uniform 
normal state into general, modulated superconducting state. 
For analytic analysis and to make use of the fewest number of plane waves in 
the OP we consider perfectly specular film surfaces 
with boundary condition 
$f(x,\pm D/2, \ul\hvp; \vare_m) = f(x,\pm D/2, \hvp; \vare_m)$, 
connecting the incoming, $\hvp$, and mirror-reflected, 
$\ul\hvp = \hvp-2\hat{\bf y} (\hat{\bf y} \hvp)$, 
trajectories at surfaces $y=\pm D/2$. 
From the self-consistency it follows that  
the OP at the surfaces is zero due to $d$-wave symmetry. 
This ensures the OP form  
$\Delta(\vR) = \sum_{q_x} \Delta_{q_x} e^{iq_x x}\, \cos \, \pi y/D$, 
and thus fixes the $y$-wave number $q_y = Q_y \equiv \pi/D$.
Note, that fixed $Q_y$ in $\eta_\vq = (v_x q_x + v_y Q_y)/2$ (Eqs.~\ref{eq:fe}) 
plays the same role as the external magnetic field in 
$\eta_{\vq B} = \vv\cdot \vq/2 + \mu B$ that enters similar equations describing 
instability into FFLO phase in Pauli-limited superconductors.\cite{sam97}

\paragraph{Phase diagram of superconducting state in a film.} 
We first find the boundary of the superconducting phase, 
and thus look for the largest $Q^*_y(T)$ 
(narrowest strip $D^*(T)$), where superconductivity first appears. 
This instability is given by condition $I(T, q_x, Q^*_y) = 0$, 
and its solution, $Q^*_y(T, q_x)$, 
is shown in Fig.~\ref{fig:PDQ}a for various $q_x$ states. 
The $q_x=0$ state, uniform along the film, 
at low temperature is preempted by states with finite modulation $q_x$. 
For any given $T$ we determine the optimal $q_x$ (denoted by $Q_x(T)$) that 
gives the largest $Q^*_y$ (see Fig.~\ref{fig:PDQ}b); 
and find that $x$-modulated states are possible below $T^* \sim 0.43 \,T_c$.  

Next, we determine the relative stability of various non-uniform states 
near this transition. 
For each instability mode $\vQ(T) \equiv (Q_x,Q_y)$ 
there are three other degenerate modes 
obtained from this one by reflection of $x$ and/or $y$ coordinates. 
We consider two principal states obtained by different combinations  
of these modes: 
(i) with two opposite $q_x$-components, corresponding to the amplitude 
oscillations of the OP, which includes all four degenerate states, 
$\vq_{1,2,3,4} = \{ (Q_x,Q_y) \,,\, (Q_x,-Q_y) \,,\, (-Q_x,-Q_y)  \,,\, (-Q_x,Q_y) \}$; 
and 
(ii) with one $q_x$-component, that gives 
a modulation of the OP phase along the film and 
therefore a superflow, 
$\vq_{1,2} = \{ (Q_x,Q_y) \,,\, (Q_x,-Q_y) \}$. 

Near the second-order transition, 
\begin{widetext}
\begin{subequations}
\bea
\Del F[Q_x,-Q_x] = -\frac{2 \; I^2(T,\vQ)}{ K_1+2 K_{12} + 2 K_{13} + 2 K_{14} +2 K_{1234} }
&\mbox{for}&
\Delta(\vR) = \Delta_1 (e^{i Q_x x} + e^{-i Q_x x}) \cos Q_y y  \,,
\label{eq:FEa}
\\
\Del F[Q_x] = -\frac{2 \; I^2(T,\vQ)}{2(K_1+2 K_{12}) } 
\quad &\mbox{for}& \quad
\Delta(\vR) = \Delta_2 e^{i Q_x x} \cos Q_y y  \,.
\label{eq:FEb}
\eea
\label{eq:FE}
\end{subequations}
\end{widetext}
Here we use notation, $K_1 = K(\vq_1,\vq_1,\vq_1,\vq_1)$, 
$K_{1234} = K(\vq_1,\vq_2,\vq_3,\vq_4)$, and for  
pairs of wave vectors, $K_{ij} = K(\vq_i,\vq_j,\vq_i,\vq_j)$. 
At low temperatures we calculate these coefficients analytically and 
find that $K_{13}$ term, corresponding to the two opposite wave
vectors $\pm \vQ$ ($\vQ \nparallel node$), diverges as $1/T$.  
This makes the current-carrying state with spontaneously 
broken time-reversal symmetry be the lowest in 
energy.\cite{footnote}
Numerical evaluation of all $K$'s 
shows that the state with a current has lowest energy at all $T<T^*$. 
We also find that transition from normal to superconducting state 
is always second order 
in the film, to be contrasted with bulk Pauli-limited superconductors, 
where $\Delta^4$-term coefficient becomes negative and 
transition becomes first order at low temperatures.\cite{sar63,mak64a}
In a film, even for $q_x=0$, we have a wavevector pair $\pm Q_y \ne 0$ 
and $K_{12}>0$, which guarantees positive sign of $K_1+2 K_{12}$ in 
Eq.~(\ref{eq:FEb}), even though $K_1<0$ at low temperatures. 

\begin{figure}[t]
\centerline{\includegraphics[width=0.85\linewidth]{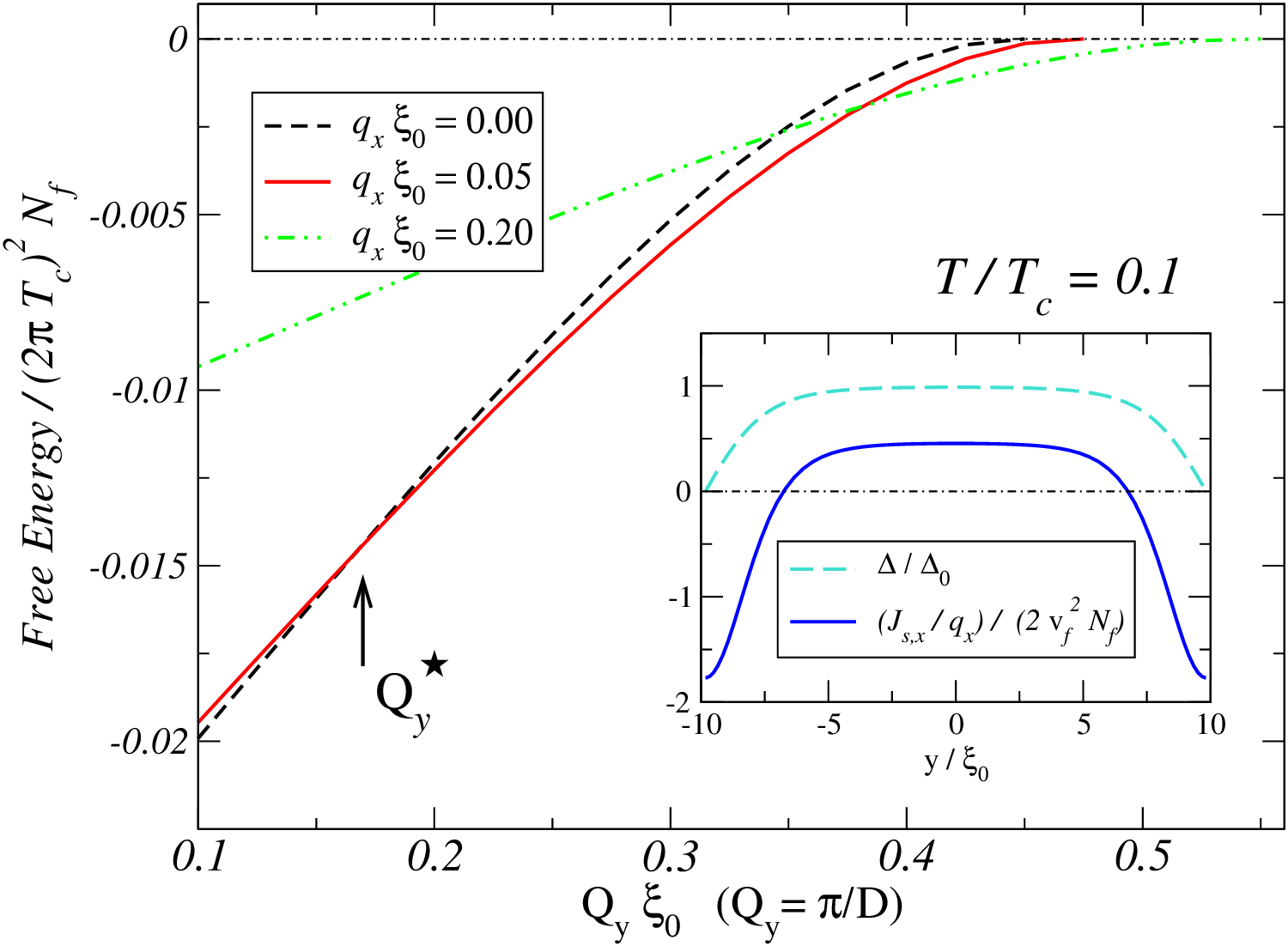}}
\caption{(Color online) 
Free energy density of a $d$-wave superconductor confined to a film, 
as a function of the inverse thickness for $T/T_c=0.1$. 
In films with $D < D^\star = \pi/ Q_y^\star$ the states 
with superflow have lower energy than the $q_x=0$ state. 
The inset shows self-consistently calculated order parameter 
profile, $\Delta(y)$, and the current density, $J_{s,x}(y)$, 
for small $q_x$ and $D \lesssim D^\star$. 
The current has anomalous (paramagnetic) contributions near the edges 
due to Andreev bound states, and 
$Q_y^\star$ marks the point where the 
average superfluid density in the film, 
$\rho_s \equiv \lim\limits_{q_x\to 0} J^D_{s,x}/q_x$, vanishes. 
}
\label{fig:curFE}
\end{figure}

\begin{figure}[t]
\centerline{\includegraphics[width=0.85\linewidth]{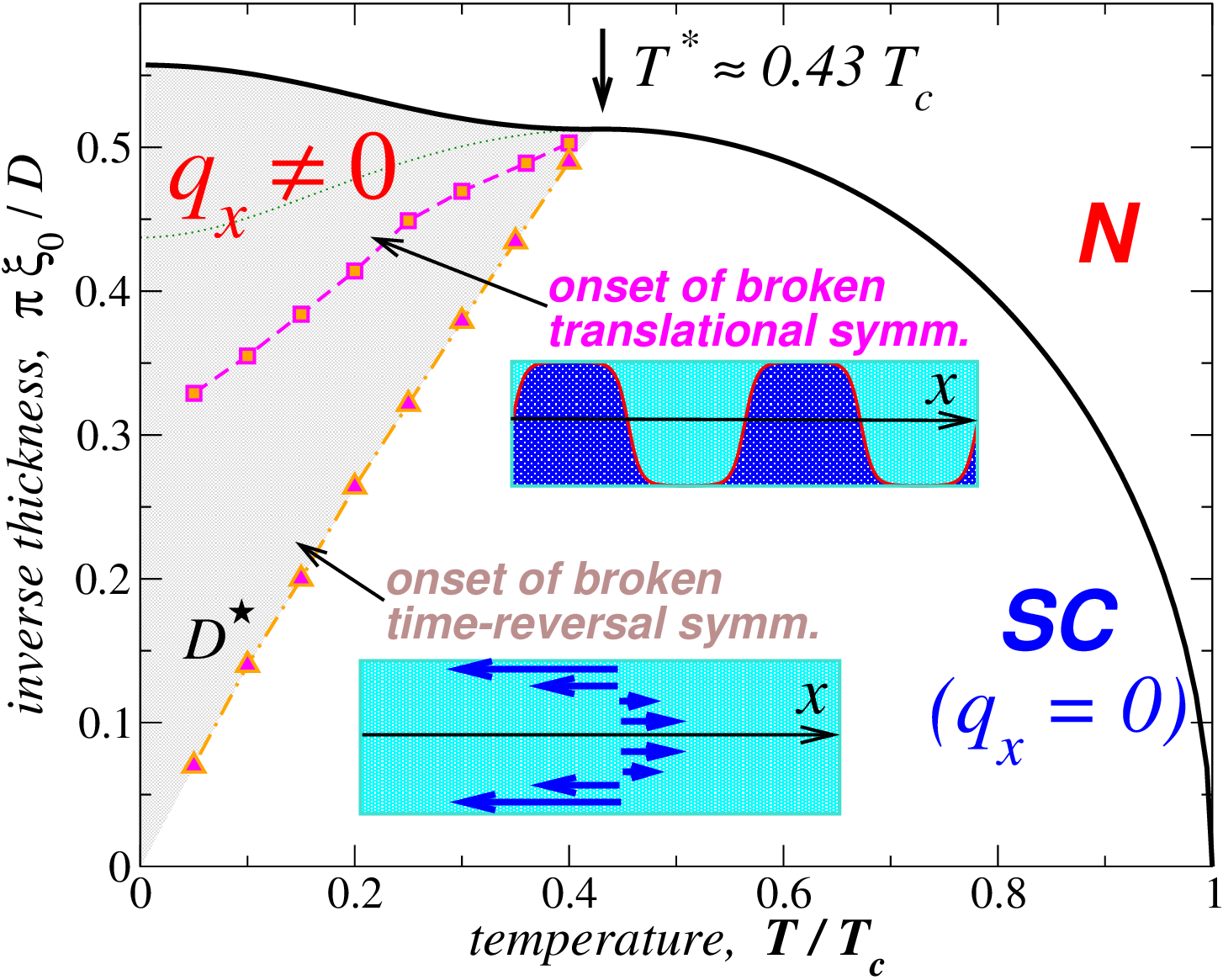}}
\caption{(Color online)
The phase diagram of a $d$-wave superconductor in a film. 
The shaded region marks the ground state with flowing current. 
Suppression of this state would open access to  
the state with broken translational symmetry. 
}
\label{fig:PD}
\end{figure}

To fully describe the phase diagram and elucidate the structure 
of the new phases, we determine the transition between the new phases 
and the $q_x=0$ `uniform' condensate deep inside the SC state. 
We start with the state that breaks translational symmetry and forms 
periodic modulations of $|\Delta(\vR)|$ in the film's plane. 
In this case the general form of the OP is similar 
to one in the FFLO problem.\cite{bur94,vor05b} 
Near the normal state instability 
$\Delta(x,y) \sim \cos Q_x x \: \cos Q_y y $, 
but becomes less harmonic and more domain-like 
structured along $x$, as we increase film thickness ($Q_y$ decreases), 
The distance between neighboring domains 
grows and at some critical $Q^{crit}_y$ the last domain wall, 
that separates two degenerate states at $x=\pm\infty$ 
with opposite OP profiles $\pm\Delta(y)$, 
disappears (square-symbol line in Fig.~\ref{fig:PD}). 
Note, that this transition occurs below unphysical 
$q_x=0$ line (thin dots in Fig.~\ref{fig:PD}), 
as compared with the FFLO problem\cite{bur94,vor05b}:  
there is no first order transition above $q_x=0$ line in this case -  
thus the inhomogeneous phase must cover the entire thermodynamically 
unstable region. 
Energetically, the state with the modulated amplitude of the OP 
gains energy compared with the energy of `uniform' state 
from the reduction of pairbreaking at the domain walls and 
the redistribution of the Andreev bound states.\cite{vor07a} 

Next, we consider state with phase modulation (superflow), 
$\Delta(\vR) = \Delta \, \cos Q_y y \, e^{iQ_x x}$, 
that carries supercurrent, 
$J_{s,x}(y) = 2 N_f \, T\sum_{\vare_m} \, 
\langle v_x(\hvp) \, g(y, \hvp; \vare_m) \rangle$. 
We write down a more general form of the OP, 
$\Delta(x,y) = \Delta_{q_x}(y) \exp(i q_x x)$, 
and self-consistently determine its amplitude profile and 
the associated free energy density\cite{vor03c}, 
\be
\Omega^D(q_x) = \int^{D/2}_{-D/2} {dy \over D} \; \Del F^D(q_x, y) \,,
\ee 
as a function of $q_x$ and the film thickness, $D$.  
Key features of this calculation are shown in Fig.~\ref{fig:curFE}. 
In the main panel we plot $\Omega^D$ as a function of inverse 
film thickness for several $q_x$. 
States with finite superflow are stabilized in films thinner 
than $D^\star(T) = \pi / Q_y^\star(T)$. 
Moreover, their stability region extends beyond that of $q_x=0$ state. 
Since finite $q_x$ induces supercurrent, it can be the ground state 
only when the total current in the film disappears  
$J^D_{s,x} \equiv \rho^D_s(q_x) q_x = 
\partial\Omega^D(q_x)/\partial q_x = 0$.\cite{kus99}
Upper critical width $D^\star$ is determined by vanishing 
average superfluid density, $\rho^{D^\star}_s(0)=0$. 
These conditions are possible to satisfy 
due to backflowing anomalous 
surface currents carried by Andreev bound states 
(inset of Fig.~\ref{fig:curFE}).\cite{fog97a,wal98,hig97}

The complete phase diagram is presented in Fig.~\ref{fig:PD}. 
Below $T^* \sim 0.43 \, T_c$, 
two new ground states are possible in superconducting films.  
The state with spontaneous current and 
broken time-reversal symmetry takes a large part of the phase space. 
Under considered conditions the state with modulated amplitude 
of the OP lies inside the stability region of the 
current-carrying state and is not realized. 
However the relative energies of the two states may be affected, 
e.g. by surface roughness. 
Also note, within numerical precision the lower instability line, 
$D^\star(T)$, is nearly straight and extrapolates to the origin. 

\begin{figure}[t]
\centerline{\includegraphics[width=0.85\linewidth]{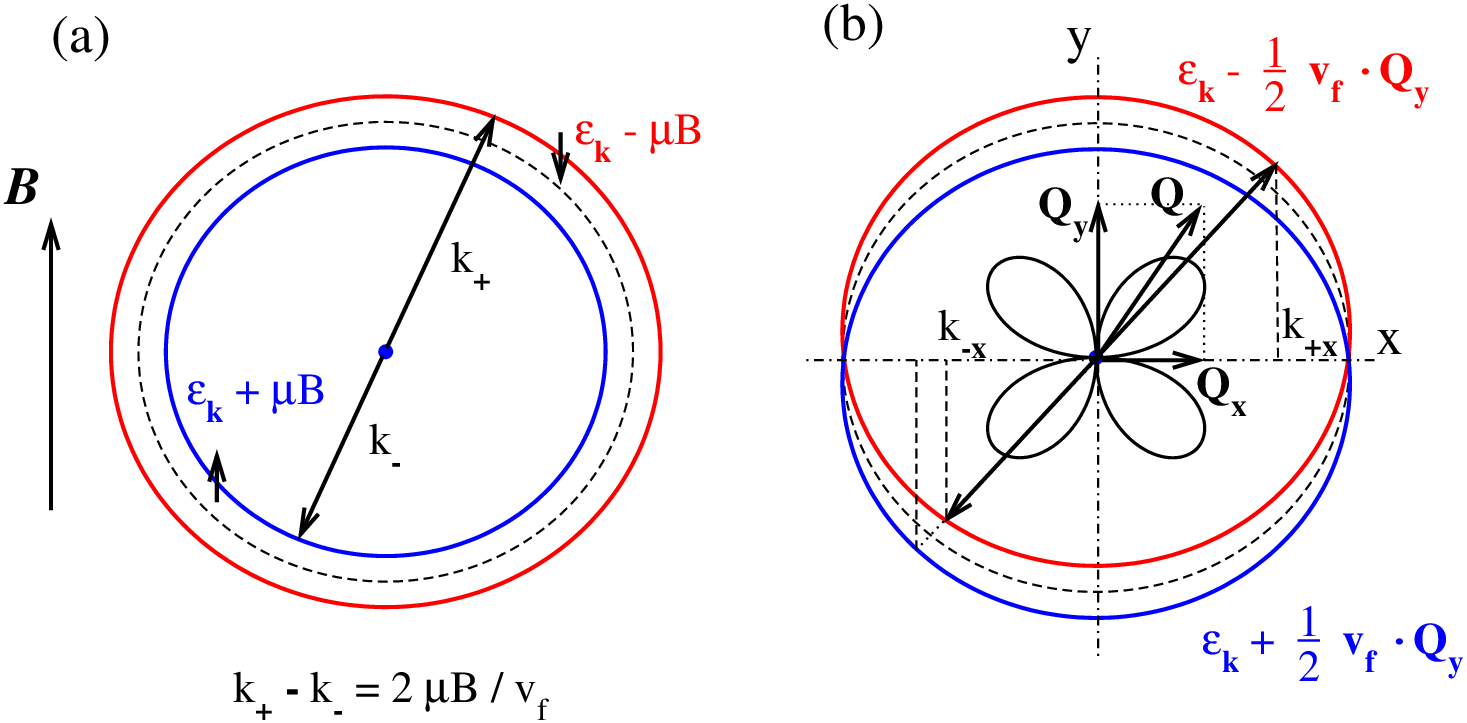}}
\caption{(Color online) 
(a) inhomogeneous state of FFLO type, when the pair-forming states 
with opposite spins belong to different Fermi surfaces, 
split by magnetic field; 
(b) state with spontaneous current along $x$. 
Fermi surface is shifted by fixed center-of-mass flow $\pm Q_y$. 
Paired states can have additional $x$-momentum $Q_x$ 
to minimize pairbreaking. 
}
\label{fig:P+P-}
\end{figure}

Qualitative picture for appearance of the longitudinal 
modulations of the OP in the film is illustrated 
by analogy with the FFLO state in Fig.~\ref{fig:P+P-}. 
In the FFLO state modulations of the OP arise to minimize 
pairbreaking caused by pairing of electrons 
across Zeeman-split Fermi surfaces, 
$\vare_\vk \pm \mu B$, Fig.~\ref{fig:P+P-}(a). 
In the strongly inhomogeneous state in the film, 
the role of magnetic field is played by the fixed confining 
wave vector $Q_y$ that produces superflow across the film and shifts the 
Fermi surface vertically, $\vare_\vk \pm v_y Q_y /2$, Fig.~\ref{fig:P+P-}(b). 
Pairs are formed with additional shift $Q_x$ (so that $\vQ \nparallel node$) 
to minimize pairbreaking by the superflow, c.f. Ref.~\onlinecite{doh06}.

Finally, we remark on the observability of these states. 
In superconductors diamagnetic coupling to 
(self-induced) magnetic field will modify the phase digram, Fig.~\ref{fig:PD}.  
Previous studies of semi-infinite ($1/D\to0$) system\cite{hig97,bar00a,lof00} 
show that spontaneous surface current appears already at finite temperature 
and this indicates that the phase space for $q_x \ne 0$ state may be enlarged. 
Also, modulated states in confined geometry are less sensitive 
to the disorder and would persist much longer in dirty samples compared with 
surface states.\cite{bar00a}
In fact, we find that complete suppression of $T^* \to 0$ by impurities 
happens only when the corresponding suppression of  $T_c$ is 60\% 
(mean free path $\ell \sim 5 \xi_0$). 
It is also reasonable to expect that as long as the OP 
has significant gradient across the film, the new superconducting phases 
are only slightly affected by surface roughness. 

\paragraph{Conclusions.} 
We have studied behavior of a $d$-wave superconductor in a film geometry. 
We find that large gradients of the order parameter across the film, 
can be - quite counterintuitively -  
`relieved' by producing additional modulation along the film. 
This means that in confined geometry this pairing system 
may undergo a transition into a state with broken translational or even
time-reversal symmetry. 
Similar behavior occurs in a completely different pairing state 
(triplet, $p$-wave) in \He\ films\cite{vor07a} and we suggest that 
these states are common to unconventional pairing systems, 
although their exact nature may depend on the structure of the OP. 
The new ground states are more robust in film or wire geometry than in 
bulk or semi-infinite systems 
and should be observable in superconductors and 
neutral superfluid \He\ in confinement. 

\paragraph{Acknowledgements.} 
I thank J.~Sauls, I.~Vekhter and P.~Adams 
for valuable discussions and suggestions on the manuscript. 
This work was supported by the Board of Regents of Louisiana. 


\begin{thebibliography}{28}
\bibitem{footnote} 
Note, that in Eqs.~(\ref{eq:eil}) and (\ref{eq:fe}) we omitted 
contributions from flowing electric currents and magnetic field, 
that would add to the fourth-order term in the free energy. 
This is done to simplify the analysis and emphasize that 
in non-uniform superconductors the state with a superflow 
may exist without the vector potential.
\end{thebibliography}


\end{document}